\documentclass{PoS}
\usepackage{slashed}
\usepackage{multirow,tabularx}

\newcommand{\mg}{{\sc MadGolem\,}}

\title{MadGolem: automating NLO calculations for New Physics}

\ShortTitle{MadGolem: automating NLO calculations for New Physics}

\author{\speaker{David L\'opez-Val},  Dorival Gon\c{c}alves-Netto, Tilman Plehn
         \\
        Institut f\"ur Theoretische Physik, Universit\"at Heidelberg\\
        E-mails: \email{lopez@thphys.uni-heidelberg.de} \\ \email{netto@thphys.uni-heidelberg.de} \\ \email{plehn@thphys.uni-heidelberg.de}}

\author{Kentarou Mawatari \\
       Theoretische Natuurkunde and IIHE/ELEM, Vrije Universiteit Brussel, Belgium \\
       International Solvay Institutes, Brussels, Belgium \\
       E-mail: \email{kentarou.mawatari@vub.ac.be}}
\author{Ioan Wigmore \\
       SUPA, School of Physics \& Astronomy, The University of Edinburgh, UK \\
       E-mail: \email{i.t.wigmore@ed.ac.uk}}
\abstract{With the LHC close to complete its 8 TeV run,
the experimental searches have already started to probe the vast beyond-the-standard Model scenery.
Providing next-to-leading order (NLO)
predictions for the major
new physics discovery channels
is therefore a most pressing request to
particle phenomenologists these days. 
\mg is a new computational tool that automates NLO calculations
of generic $2\to2$ new physics processes in the {\sc MadGraph/Golem} framework.
In this contribution we concisely describe the structure and performance of the code,
with particular focus on the 
generation of the renormalized one-loop amplitudes and
the automatized subtraction of infrared and on-shell divergences.
We briefly survey the many dedicated tests of all these aspects and outline
some applications to LHC phenomenology.}

\FullConference{Loops and Legs in Quantum Field Theory - 11th DESY Workshop on Elementary Particle Physics,\\
		April 15-20, 2012\\
		Wernigerode, Germany}

\begin{document}

\section{Introduction}

As a result of an outstanding performance, the LHC has already
delivered around 15 fb$^{-1}$ of data at the ATLAS and CMS detectors, 
setting out on the quest for signatures of new physics. Direct searches
rely on the pairwise production of novel heavy states or, alternatively, on their associated production
along with SM degrees of freedom. 
Just to mention one example, the studies conducted so far have already
enabled to constrain the
phenomenologically viable parameter space of the Minimal Supersymmetric Standard Model (MSSM),
mainly upon analyses based on jet production from squark and gluino
decays plus missing energy.
Accurate predictions, including next-to-leading order (NLO) QCD corrections,
are thus instrumental at this stage. They allow to reduce the theoretical uncertainty,
as they render more stable results with respect to the (unphysical) renormalization and factorization scale choices;
on the other hand, they provide suitable total rates to normalize the event samples simulated by standard Monte Carlo (MC) generators.
In this context, the development of dedicated tools that automate this type of computations ranks very
high in the phenomenologists' wishlist. The community
has witnessed an impressive thrust of activity in the recent years and achieved very significant milestones, cf. e.g. 
\cite{sqn1,sgluon,auto}. 

\smallskip{}
\mg \cite{sqn1,sgluon} 
is conceived as a highly modular, independent add-on to the major MC generator {\sc MadGraph} \cite{madgraph}.
It implements an automated framework in which to compute total cross sections and distributions,
including QCD quantum effects to NLO accuracy. \mg is mainly tailored to describe the production of heavy particle pairs
within theories beyond the SM. The tool is currently undergoing a final testing phase, prior to its
public release, and is
meant to be of interest for model-builders, phenomenologists and fundamentally for the LHC experimental community.

\section{Code structure}

\begin{figure}[t]
 \begin{center}
  \includegraphics[width=0.8\textwidth]{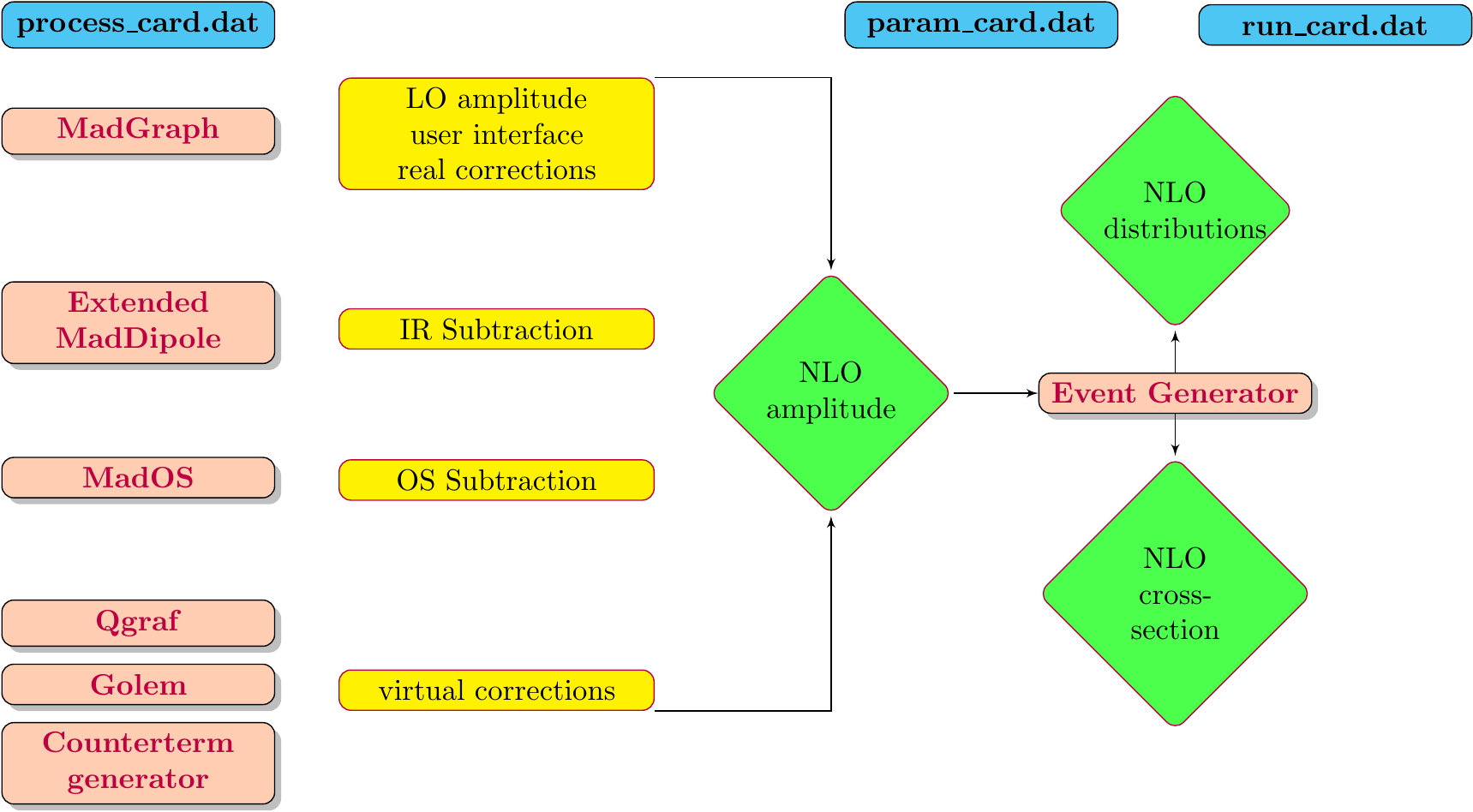}
\caption{\footnotesize{Modular structure of {\sc MadGolem}.}}
\label{fig:flowchart}
 \end{center}
\end{figure}

A schematic picture of the flowchart and the inner architecture of {\sc MadGolem}
we display in Figure~\ref{fig:flowchart}. 
The tree-level matrix elements (i.e. the leading order $2\to 2$ ones as well as the $2\to3$ contributions from real
emission) we obtain via {\sc MadGraph 4.5} \cite{madgraph}. The one-loop 
Feynman diagrams we generate via {\sc Qgraf} \cite{qgraf}. The latter we translate
and further process through a dedicated chain of
{\sc Perl}, {\sc Form} and {\sc Maple} routines that handle the corresponding color, helicity 
and tensor structures. 
Spinors we manipulate algebraically resorting to spinor-helicity methods \cite{spinor}, while
we deal with  
color algebra by means of a conventional color flow decomposition technique \cite{color}. 
Tensor structures we decompose
via a modified Passarino-Veltman scheme in the framework of {\sc Golem} \cite{golem}.
Both ultraviolet (UV) and infrared (IR) singularities
from the loop integrals we regulate employing the standard 'tHooft-Veltman dimensional regularization
prescription in $n = 4-2\epsilon$ dimensions. 
In order to handle these one-loop corrections we stick to a fully analytical, Feynman-diagrammatic approach.
The explicit analytical
form of the one-loop amplitudes we keep accessible at all the stages throughout the entire calculation. The
final expression we cast in terms of partial amplitudes, each of them consisting of i) 
one coefficient, which depends on the coupling constants, masses and kinematic invariants;
and ii) a basis of fundamental color and helicity structures, as well as of one-loop integrals.
The latter we evaluate numerically with the aid of {\sc OneLoop} \cite{oneloop}. 

\smallskip{}
A major building-block of \mg concerns the automated handling of the different divergent
contributions, which are ubiquitous in higher order calculations. UV divergences, on the
one hand, we renormalize by means of corresponding UV counterterms, which we generate
via {\sc MadGraph 4.5}\cite{madgraph} alongside the tree-level amplitudes.
These counterterms we write in terms of $\mathcal{O}(\alpha_s)$, model-dependent 
2-point functions which we supply as a separate library. We use the $\overline{\rm MS}$ scheme with decoupled
heavy colored states~\cite{decoupling} for the renormalization of the strong coupling constant $\alpha_s$, while 
the particle masses we renormalize on-shell.
If applicable, Supersymmetry-restoring counterterms are conveniently included at this point \cite{susyrestore}.
Catani-Seymour (CS) dipoles we introduce to subtract the collinear and soft
singularities \cite{cs}. We implement them as a generalization to the {\sc MadDipole} 
package\cite{maddipole}, including the novel massive dipoles needed to cope
with the IR structures of the non-standard heavy colored particles
(e.g. squarks, gluinos or sgluons, among others). We retain
the explicit dependence
on the FKS-like $\alpha$ phase space 
cutoff \cite{alpha}. In so doing we explicitly track down the separation
of the soft and collinear 
regions that are subsumed into the integrated dipoles -- and so into part of the virtual corrections --
and those which constitute the genuine contribution to the $2 \to3$ real emission terms.
Additionally, on-shell (OS) divergences may occur if the produced heavy colored particles give
rise to light-quark jets, as part of the real emission corrections at NLO. These situations lead 
to non-integrable phase space singularities and induce a potential double counting,
if effectively treated as part of the NLO effects. We adopt the {\sc Prospino} scheme 
\cite{prospino} to subtract these singularities locally and sidestep the 
mentioned double counting, a strategy that preserves gauge invariance
and spin correlations.

\smallskip{}
Finally, we interface all these different modules and translate the resulting analytical
output into Fortran code, which we can further run numerically to
obtain the NLO total rates and distributions. A user-friendly interface is exported
from {\sc MadGraph}. The model parameters, collider setup and numerical
arrangements can be specified by the user via a set of input cards, 
again following the familiar
{\sc Madgraph} environment. The structure of the resulting Fortran code, in terms
of independent subprocesses for the leading-order (LO), virtual and real corrections, 
for each of the different partonic subchannels, makes it particularly
suitable for parallelization. 

\smallskip \mg is mainly oriented towards 
beyond the Standard Model (BSM) physics. 
As compared to alternative codes, such as the case
of {\sc Prospino} -- specifically tailored for pair-production processes 
within the MSSM -- \mg
offers a flexible, adaptable and fully automatized platform. It
enables a broad coverage of the new physics scenery, as it handles
genuine BSM structures such as Majorana fermions, new IR and OS singularities
stemming from heavy colored states, and effective interactions -- 
the latter parameterized by
higher dimensional operators. Currently, \mg fully supports
NLO calculations within i) the SM; ii) the MSSM; iii) a number of generic new physics
realizations featuring new heavy states with different color charges and
spin representations (e.g. sgluons and leptogluons, among others); iv) generic extensions
of the SM with higher dimensional operators
(e.g. for the study of monotop signatures from dimension six operators).

\smallskip{}
The complexity of some processes, in particular those involving fermion fields
with large color representations, demands for dedicated coding strategies
to achieve a satisfactory performance of the tool at the different 
stages, in particular in terms of processing and running times. 
These strategies include, for instance, loop filtering; grouping of topologically equivalent
diagrams; and the intensive use of dynamically-linked libraries and multithreat processing.

\section{Validation strategies and applications to phenomenology}

\begin{figure}[t!]
\begin{minipage}{0.3\linewidth}

\begin{tabular}{c||l|l|c} \hline
 & \multicolumn{3}{c}{$\sqrt{S} = 7$ TeV \,}  \\ \hline
 $m_{G}$ [GeV]  & $\sigma^{\rm LO} [{\rm pb}]$ & $\sigma^{\rm NLO} [{\rm pb}]$ & $K$   \\ \hline\hline
 200 & $1.40 \times 10^2$ & $2.26\times 10^2$ & 1.61   \\
 350 & $4.83\times 10^{0}$ & $8.21\times 10^{0}$ & 1.70 \\
 500 & $4.05\times10^{-1}$ & $7.32\times10^{-1}$ & 1.81 \\
 750 & $1.48\times10^{-2}$ & $3.01\times10^{-2}$ & 2.03 \\
 1000 & $8.60\times 10^{-4}$ & $2.00\times 10^{-3}$ & 2.33 \\ \hline
\end{tabular}

\end{minipage} 
\hspace{3.5cm}
\begin{minipage}{0.54\linewidth}
\includegraphics[width=0.9\textwidth]{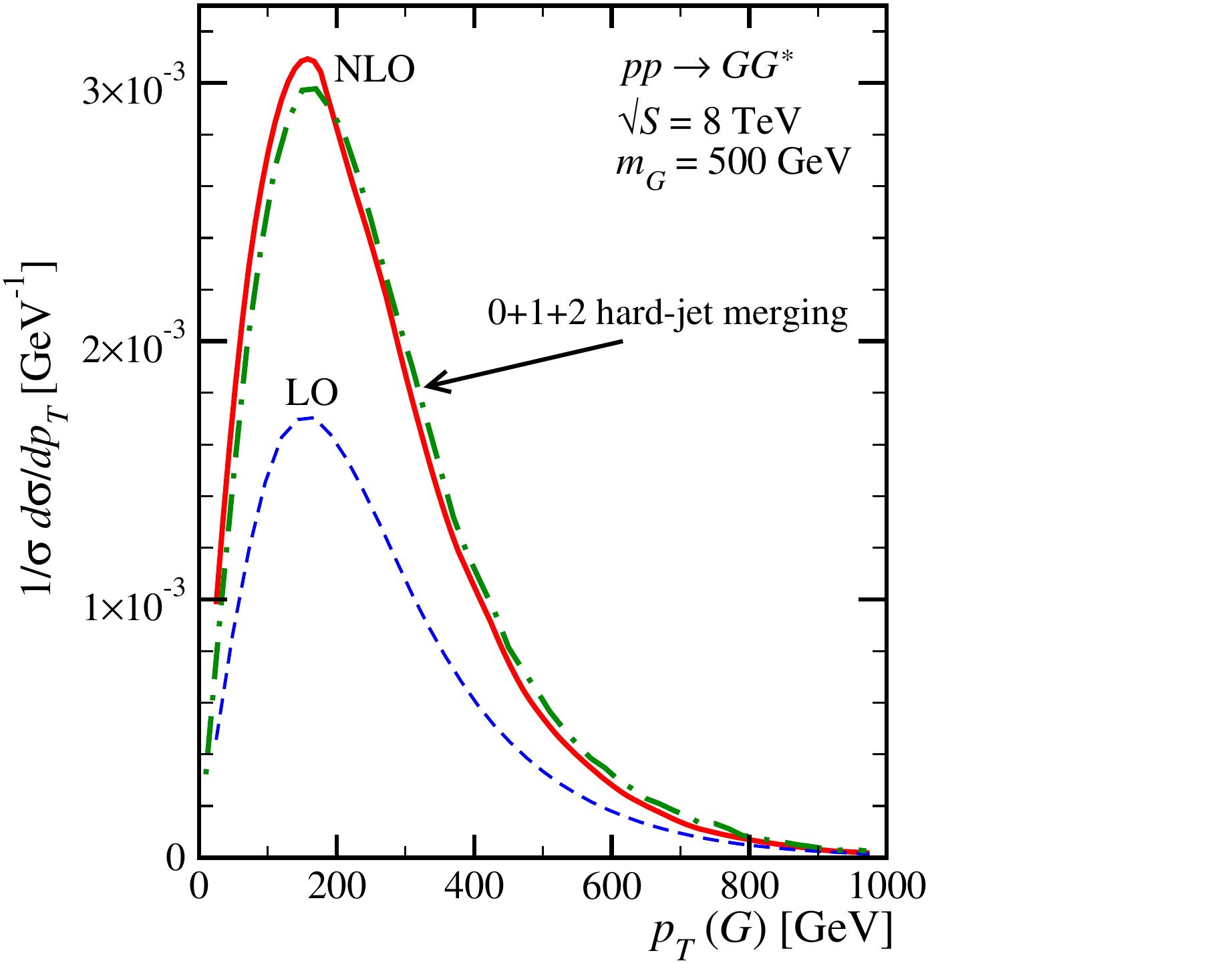}
\end{minipage}
\caption{Example of a phenomenological analysis carried out within {\sc MadGolem}. We consider the production
of sgluon pairs at the LHC \cite{sgluon}. In the left panel we show the total LO and NLO rates
for different sgluon masses, alongside with the associated $K$-factors. In the right panel
we display the LO and NLO distributions as a function of the sgluon $p_T$, and compare them to 
the results from MLM jet merging with (0+1+2) hard jets.}
\label{fig:sample}
\end{figure}

Exhaustive cross-checks have been performed to ensure the reliability of our tool.
The total NLO rates and corresponding $K$ factors
have been computed with \mg for 
a fairly large number of processes 
both within the SM and the MSSM, covering all representative possibilities
of spin/color representations, interactions and topologies. The cancellation of the
UV, IR and OS divergences, as well as the gauge invariance 
of the overall result, has been explicitly confirmed (in all cases numerically, 
and also analytically for some specific ones). The finite parts
of the renormalized one-loop amplitudes we have contrasted against
independent calculations performed with
{\sc FeynArts}, {\sc FormCalc} and {\sc LoopTools} \cite{feynarts}. 
Dedicated attention has been
devoted to the numerical stability and convergence of the results, 
ensuring a robust implementation of the CS dipoles
and the OS subtraction scheme. The specific behavior
of the dipoles, as well
as of the OS subtraction terms nearby the singular regions, 
has been carefully studied -- including e.g.
the aforementioned $\alpha$-parameter to ascertain the independence of 
the subtraction procedure from the arbitrary phase space regulator in use. 
Finally, and whenever available, we have compared the \mg outcomes
to independent results presented in the literature.
For the total NLO rates, in particular, we have explicitly confirmed our agreement
with {\sc Prospino} \cite{prospino_chargino} for
the main pair production channels within the MSSM, including the processes
mediated by (SUSY)QCD (viz. $pp \to \tilde{q}\tilde{q}$, $\tilde{q}\tilde{q}^*$, $\tilde{q}\tilde{g}$
and $\tilde{g}\tilde{g}$) and SUSY-EW interactions ($pp \to \tilde{\chi}\tilde{\chi}, \tilde{l}\tilde{l}^*$). 
For the distributions, in turn, we have cross-checked our results with 
 {\sc MadFKS} \cite{madfks}.

\bigskip{}
Letting aside this extensive cross-checking program, we have
also started to exploit all the \mg capabilities
to conduct original phenomenology studies. 
On the one hand, we have considered the associated squark/neutralino production 
$pp \to \tilde{q}\tilde{\chi^0}$, which leads to a characteristic monojet $+\slashed{E_T}$ signature \cite{sqn1}; 
likewise, we have addressed the pair production of new exotic states with higher color
representations, in particular the case of scalar $SU(3)_C$ adjoints -- the so-called
sgluons~\cite{sgluon}. A number of further applications are also underway, in particular 
a comprehensive study of the pairwise production of squarks and gluinos \cite{mainpaper},
which spells out the improved capabilities of \mg with respect to 
the presently available tools \cite{prospino}.

\smallskip{}
All these analyses thus 
qualify as pioneering examples of totally automated NLO calculations 
in a wide variety of new physics scenarios.
\mg enables to perform very detailed phenomenological studies
of pair production processes beyond the SM, including
i) the calculation of the total NLO rates and corresponding $K$ factors; in the
case of the MSSM, for instance, it is possible to undertake a
comprehensive survey along the parameter space of the model \cite{mainpaper},
as no restrictions on the SUSY mass spectrum nor the couplings are
assumed beforehand -- at variance with alternative  codes; ii)
a detailed analysis of the virtual and the real NLO quantum corrections -- as the different
contributions from all the subchannels and one-loop topologies are accessible
separately; iii) the study 
of the dependence with respect to the renormalization and factorization scale choices, and
so of the theoretical uncertainties and how these nail down when comparing the NLO
predictions to the mere LO ones;
iv) and perhaps most significantly, \mg supports the generation of NLO distributions. Interestingly enough,
further comparisons to the results obtained via multi-jet merging \cite{mlm} can be easily
carried out in the {\sc MadGraph} framework. These nicely confirm that the matched samples with
extra hard jets correctly reproduce the kinematical features of colored heavy particle production,
while the total NLO rates from the fixed-order calculation can be taken as a suitable normalization
for the corresponding event samples. In Figure~\ref{fig:sample} we illustrate some of these features
for the particular case of sgluon pair production
$pp \to GG^*$ at the LHC \cite{sgluon}.

\bigskip{}
In summary, \mg completely automates the calculation of NLO QCD corrections to generic BSM pair production
processes, as well as their interface to the standard Monte Carlo generator {\sc MadGraph}.
The tool resorts to a fully analytical, Feynman-diagrammatic approach and is endowed with all
the technical machinery needed to automatically handle the UV, IR and OS singularities and to retrieve
the total NLO cross-sections and corresponding $K$-factors, as well as distributions.   
\mg represents a new step forward in this automation program which sets sights, after all, on extending
solid bridges between theory and experiments. 

\footnotesize{\paragraph{Aknowledgements}
DLV wishes to thank the organizers of the LL2012 workshop, and very particularly Prof. Tord Riemann,
for the opportunity to present our tool and for the lively atmospheare we all shared in Wernigerode.}

\providecommand{\href}[2]{#2}
\addcontentsline{toc}{section}{References}
\bibliographystyle{JHEP}

\begin{thebibliography}{99}


\bibitem{sqn1}
  T.~Binoth, D.~Gon\c{c}alves-Netto, D.~L\'opez-Val, K.~Mawatari, T.~Plehn, and I.~Wigmore,
  Phys.\ Rev.\  D {\bf 84}, 075005 (2011).
\bibitem{sgluon} D.~Gon\c{c}alves-Netto, D.~L\'opez-Val, K.~Mawatari, T.~Plehn and I.~Wigmore,
  Phys.\ Rev.\  D {\bf 85}, 114024 (2012).

\bibitem{auto} 
  G.~Cullen, N.~Greiner, G.~Heinrich, G.~Luisoni, P.~Mastrolia, G.~Ossola, T.~Reiter and F.~Tramontano,
  Eur.\ Phys.\ J.\ C {\bf 72}, 1889 (2012); 
  V.~Hirschi, R.~Frederix, S.~Frixione, M.~V.~Garzelli, F.~Maltoni and R.~Pittau,
  JHEP {\bf 1105}, 044 (2011); 
  G.~Bevilacqua, M.~Czakon, M.~V.~Garzelli, A.~van Hameren, Y.~Malamos, C.~G.~Papadopoulos, R.~Pittau and M.~Worek,
  Nucl.\ Phys.\ Proc.\ Suppl.\  {\bf 205-206}, 211 (2010).



\bibitem{madgraph} 
 J.~Alwall {\it et al.},
  JHEP {\bf 0709}, 028 (2007).
%

\bibitem{qgraf} 
 P.~Nogueira,
  J.\ Comput.\ Phys.\  {\bf 105}, 279 (1993).

\bibitem{spinor}
  Z.~Xu, D.~-H.~Zhang and L.~Chang,
  Nucl.\ Phys.\ B {\bf 291}, 392 (1987).

\bibitem{color} Cf. e.g. 
  L.~J.~Dixon,
  ``Calculating scattering amplitudes efficiently,''
   In *Boulder 1995, QCD and beyond* 539-582
  [hep-ph/9601359], and references therein.


\bibitem{golem}
 G.~Cullen, N.~Greiner, A.~Guffanti, J.~-P.~Guillet, G.~Heinrich, S.~Karg, N.~Kauer and T.~Kleinschmidt {\it et al.},
  Nucl.\ Phys.\ Proc.\ Suppl.\  {\bf 205-206}, 67 (2010);
 T.~Binoth, J.~P.~Guillet, G.~Heinrich, E.~Pilon, and T.~Reiter,
  Comput.\ Phys.\ Commun.\  {\bf 180}, 2317 (2009);
 N.~Greiner, A.~Guffanti, T.~Reiter and J.~Reuter,
  Phys.\ Rev.\ Lett.\  {\bf 107}, 102002 (2011); 
 G.~Cullen, J.~P.~.Guillet, G.~Heinrich, T.~Kleinschmidt, E.~Pilon, T.~Reiter and M.~Rodgers,
  Comput.\ Phys.\ Commun.\  {\bf 182}, 2276 (2011).


\bibitem{oneloop}
  A.~van Hameren,
  Comput.\ Phys.\ Commun.\  {\bf 182}, 2427 (2011).

\bibitem{decoupling}
 J.~C.~Collins, F.~Wilczek, and A.~Zee,
  Phys.\ Rev.\  D {\bf 18}, 242 (1978);
 S.~Berge, W.~Hollik, W.~M.~M\"osle, and D.~Wackeroth,
  Phys.\ Rev.\  D {\bf 76}, 034016 (2007).

\bibitem{susyrestore}
 S.~P.~Martin and M.~T.~Vaughn,
  Phys.\ Lett.\  B {\bf 318}, 331 (1993).

\bibitem{cs}
  S.~Catani and M.~H.~Seymour,
  Phys.\ Lett.\ B {\bf 378}, 287 (1996);
 S.~Catani and M.~H.~Seymour,
  Nucl.\ Phys.\ B {\bf 485}, 291 (1997)
  [Erratum-ibid.\ B {\bf 510}, 503 (1998)]; 
 S.~Catani, S.~Dittmaier, M.~H.~Seymour and Z.~Trocsanyi,
  Nucl.\ Phys.\ B {\bf 627}, 189 (2002).

\bibitem{maddipole} 
 R.~Frederix, T.~Gehrmann, and N.~Greiner,
  JHEP {\bf 0809}, 122 (2008);
  and
  JHEP {\bf 1006}, 086 (2010).

\bibitem{alpha} 
 S.~Frixione, Z.~Kunszt, and A.~Signer,
  Nucl.\ Phys.\  B {\bf 467}, 399 (1996);
 Z.~Nagy and Z.~Trocsanyi,
  Phys.\ Rev.\  D {\bf 59}, 014020 (1999)
  [Erratum-ibid.\  D {\bf 62}, 099902 (2000)].

\bibitem{prospino} See e.g. 
 W.~Beenakker, R.~H\"opker, M.~Spira, and P.~M.~Zerwas,
  Phys.\ Rev.\ Lett.\  {\bf 74}, 2905 (1995);
 W.~Beenakker, R.~H\"opker, M.~Spira and P.~M.~Zerwas,
  Nucl.\ Phys.\ B {\bf 492}, 51 (1997); 
 W.~Beenakker, M.~Kr\"amer, T.~Plehn, M.~Spira and P.~M.~Zerwas,
  Nucl.\ Phys.\ B {\bf 515}, 3 (1998).

\bibitem{feynarts} 
 T.~Hahn,
  Comput.\ Phys.\ Commun.\  {\bf 140}, 418 (2001);
 T.~Hahn and M.~P\'erez-Victoria,
  Comput.\ Phys.\ Commun.\  {\bf 118}, 153 (1999);
 T.~Hahn and C.~Schappacher,
  Comput.\ Phys.\ Commun.\  {\bf 143}, 54 (2002);
 T.~Hahn and M.~Rauch,
  Nucl.\ Phys.\ Proc.\ Suppl.\  {\bf 157}, 236 (2006).

\bibitem{prospino_chargino}
 W.~Beenakker, M.~Klasen, M.~Kr\"amer, T.~Plehn, M.~Spira and P.~M.~Zerwas,
  Phys.\ Rev.\ Lett.\  {\bf 83}, 3780 (1999)
  [Erratum-ibid.\  {\bf 100}, 029901 (2008)].


\bibitem{madfks}
  R.~Frederix, S.~Frixione, F.~Maltoni and T.~Stelzer,
  JHEP {\bf 0910}, 003 (2009).

\bibitem{mainpaper} D.~Gon\c{c}alves-Netto, D.~L\'opez-Val, K.~Mawatari, T.~Plehn and I.~Wigmore,
in preparation.

\bibitem{mlm}
 M.~L.~Mangano, M.~Moretti and R.~Pittau,
  Nucl.\ Phys.\  B {\bf 632} (2002) 343.

\end{thebibliography}

\end{document}